# Variable Stars in the MACHO Collaboration[1] Database

Kem H. Cook[2]

*Institute of Geophysics and Planetary Physics, Lawrence Livermore National Laboratory, MS L-413, Livermore, CA 94550, U.S.A.*

C. Alcock[2]

*Lawrence Livermore National Laboratory, Livermore, CA 94550, U.S.A.*

R.A. Allsman, T.S. Axelrod, K.C. Freeman, B.A. Peterson, P.J. Quinn, A.W. Rodgers

*Mt. Stromlo and Siding Spring Observatories, Weston, ACT 2611, Australia*

D.P. Bennett[2], J. Reimann

*Center for Particle Astrophysics, Univ. of California, Berkeley, CA, U.S.A.*

K. Griest[2]

*Univ. of California, San Diego, CA 92093, U.S.A.*

S.L. Marshall[2]

*Univ. of California, Santa Barbara, CA 93106, U.S.A.*

M.R. Pratt[2], C.W. Stubbs[2]

*Univ. of Washington, Seattle, WA 98195, U.S.A.*

W. Sutherland[2]

*Univ. of Oxford, Oxford OX1 3RH, U.K.*

and D. Welch

*McMaster Univ., Hamilton, ON L8S 4M1, Canada*

**Abstract.** The MACHO Collaboration's search for baryonic dark matter via its gravitational microlensing signature has generated a massive

---

database of time ordered photometry of millions of stars in the LMC and the bulge of the Milky Way. The search's experimental design and capabilities are reviewed and the dark matter results are briefly noted. Preliminary analysis of the $\sim 39,000$ variable stars discovered in the LMC database is presented and examples of periodic variables are shown. A class of aperiodically variable Be stars is described which is the closest background to microlensing which has been found. Plans for future work on variable stars using the MACHO data are described.

1.  Introduction

The MACHO Collaboration is surveying fields toward the Large Magellanic Cloud (LMC), the Small Magellanic Cloud (SMC) and the bulge of the Milky Way to detect the transient brightenings of stars in those fields due to gravitational microlensing caused by intervening compact, massive objects. The primary goal of this project is to determine the contribution of baryonic dark matter in the form of massive, compact, halo objects (MACHOs) to the total mass of the Milky Way. It is straightforward to calculate that, at any given instant, about one in two million LMC stars will appear at least 34% brighter than normal due to microlensing if the dark halo is only composed of MACHOs (Paczynski 1986, Griest 1991). If the halo extends beyond the distance of the LMC, then the probability that an SMC star will be lensed is higher. There are two reasons to monitor stars toward the center of the galaxy. First, it provides a proof-of-principle for the experiment because microlensing is expected from the known, faint-end of the stellar luminosity function; second, the probability of seeing microlensing due to MACHOs is not much less than that expected toward the LMC (Griest et al. 1991)..

Microlensing occurs when a MACHO passes close to the line of sight to a more distant star. The MACHO's mass creates a gravitational lens resulting in two images of the source star. These images are separated by less than a milli-arcsecond for halo object masses less than thousands of solar masses and so the phenomenon is called microlensing. The total flux in the two images is more than in the unlensed situation, and so the source star appears brighter. Because the MACHO, the solar system and the source star are all in relative motion, the brightening is transient; because of the equivalence principle the brightening is achromatic. The brightening is expected to be symmetric because the apparent motions are constant over the time frame of the event. The event duration depends upon the motion of the MACHO relative to the line of sight, the mass of the MACHO and the relative distance of the MACHO and the source (see Paczynski 1986). The peak amplification of the source star depends only upon how close to the line of sight the MACHO approaches. For a lens at 10 kpc and a typical, relative velocity of 200 km/s, and a source in the LMC, the time that the source appears more than 1.34 times brighter than normal is $\hat{t} \sim 140\sqrt{M/\,M_\odot}$ days. Thus, a credible MACHO search requires monitoring millions of stars with reasonable photometric accuracy on time scales from hours to years. It also requires the demonstration that microlensing can be clearly distinguished from all possible sources of stellar variability.

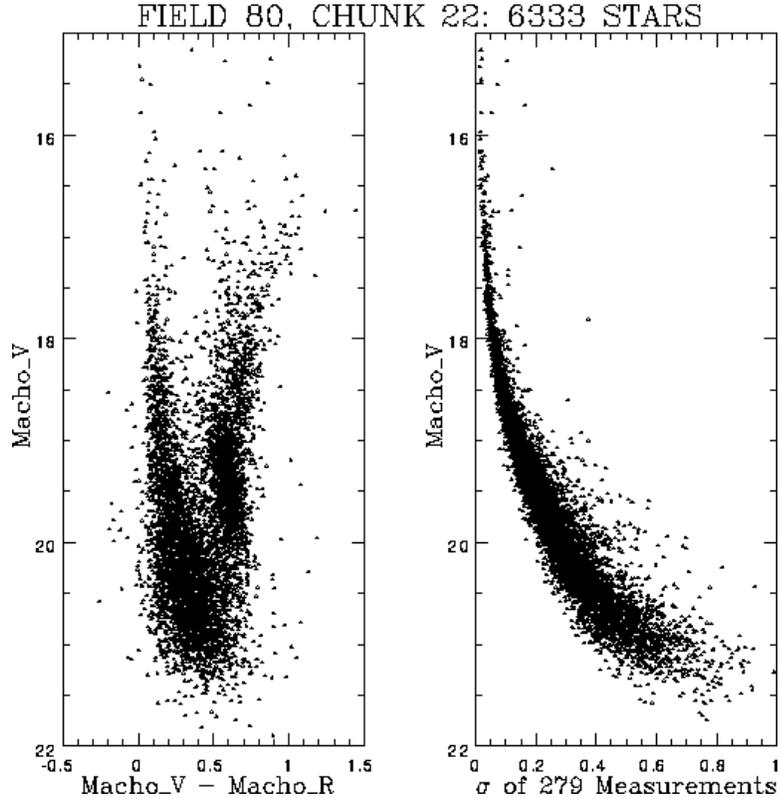

Figure 1. The left panel is a color magnitude diagram prepared from the median of 279 measurements of the stars in a 3x3 arcmin region of the LMC bar. The right panel plots the standard deviation of the time series used to calculate the median MACHO_V magnitude.

## 2. Data Collection and Analysis

The MACHO Collaboration has the dedicated use of the refurbished, 1.27m, Great Melbourne telescope for searching for microlensing. We have equipped it with a prime focus reimager-corrector with an integral dichroic beam splitter which provides a 1 degree field in two passbands simultaneously. These are each sampled with an 2x2 array of 2048x2048 Loral CCDs which are read out simultaneously via two amplifiers per CCD in about 70 seconds (Stubbs et al. 1992). These arrays produce images which cover 0.5 square degrees with 0.63 arcsec pixels.

Exposures are 300 s for the LMC and 150 s for the bulge. The telescope and data taking are automated (Alcock et al. 1992) although there is an observer present to optimize data taking and field selection during bad weather and to protect the telescope.

The observing strategy for the data taken in 1992-1993 was to observe the LMC if possible; if not, the bulge was observed. The telescope is restricted to hour angles between about ± 9 hours for the LMC because of the south support pier for the polar axle. Significant data on the SMC has only been collected since 1994. For each primary target, a set of field centers is defined so that there would be little, if any, overlap between images of adjacent fields. Each night fields are sampled such that denser fields are sampled first and progressively less dense fields are sampled later. For the LMC, this resulted in a spiral sampling centered on the bar. After sampling roughly the top 20 fields, the spiral was begun again. Images are taken whenever the weather allows. Images are bias subtracted, flat fielded and crosstalk corrected before photometry and archiving on 8mm tape. One of the best seeing images of each field is reduced to create a template of source positions and magnitudes. Routine reductions of each field are done on 3x3 arcmin chunks using about 20 bright, uncrowded stars from the template to determine the astrometric solution and define an analytic point spread function and a photometric zero point which are used to determine the brightnesses of the chunk's stars relative to these standards. The photometry code is a derivative of DoPhot (Schechter et al. 1994) which we have named Sodophot. Sodophot can photometer about $10^6$ stars per hour using a Sparc 2. The microlensing search only requires accurate relative photometry and as a result the zero points for each field and the transformation from the MACHO bands (4500-5900Å and 5900-7800Å) to standard astronomical bands have not yet been accurately determined although the data has now been collected. Current magnitude estimates are probably good to ± 0.15 mag. It should be noted that other collaborations are searching for MACHOs and finding many variable stars— in particular, see the paper by Beaulieu (1995) describing Cepheid results from the EROS experiment.

Figure 1 shows a color-magnitude diagram (CMD) for one chunk of data from an LMC bar field and the standard deviation for the light curve of each star in the CMD. This figure shows how well relative photometry can be done as a function of magnitude. The top fields from two LMC seasons and one bulge season have been reduced and analyzed for microlensing (Alcock et al. 1993, 1995a, 1995b, 1995c). The LMC data represent data on about 9 million stars with 140 to 350 epochs of data spanning 400 days and the bulge data represent 11 million stars with about 100 epochs of data spanning 150 days. The three microlensing events seen in this LMC data set suggest that we are detecting only about 20% of a standard dark halo in MACHOs of sub-stellar mass. The bulge results suggest that the structure of the inner portion of the Milky Way is not known because standard models predict much less lensing than is observed.

## 3. Periodic Variable Stars in the MACHO Database for the LMC

As part of the analysis of every source's photometry time series, a suite of statistics is calculated. For each light curve a robust $\chi^2/\text{dof}$ (where the extreme 20% of points are not used in the fit) for a fit to constant brightness is calculated and if this is larger than a (normal) $\chi^2/\text{dof}$ which can occur by chance 1% of the time, the source is considered to be variable. For sources where the psf fitting did not converge in the template frame, no analysis is done. This cut, coupled with the fact that LMC stars brighter than about V = 14 are saturated in our data, results in the exclusion of the brightest variables in the LMC from our data. It was recognized that this variable criterion selects against very long period eclipsing binaries. So, we have selected a further set of variables on the basis of their having a (non-robust) $\chi^2/\text{dof} > 3$ for downward excursions from the mean. The initial criterion yielded 38753 candidate variables out of 9 million light curves. The analysis for eclipsing binaries yielded another 10962 candidates. The candidate variable star light curves were then folded and tested for periodicity in the range of 0.1 day to the full data span which was usually about 400 days. The code used was developed by James Reimann as part of a statistics thesis at UC Berkeley (Reimann 1994) and employs the super-smoother algorithm to model the folded light curve. A measure of the goodness of fit of the model and the data is calculated. This technique basically determines whether the trial period has produced a smooth folded light curve and is fast and robust. The light curves were divided into categories on the basis of their mean magnitudes and colors and sorted on the basis of the goodness of fit parameter of the best period. Our categories include Cepheids, RR Lyrae, eclipsing binaries and red variables. This initial data set is not long enough to determine periods for many of the Long Period Variables and so we have chosen to call all variables which are redder than V−R = 0.9 red variables until we have longer time spans to fold.

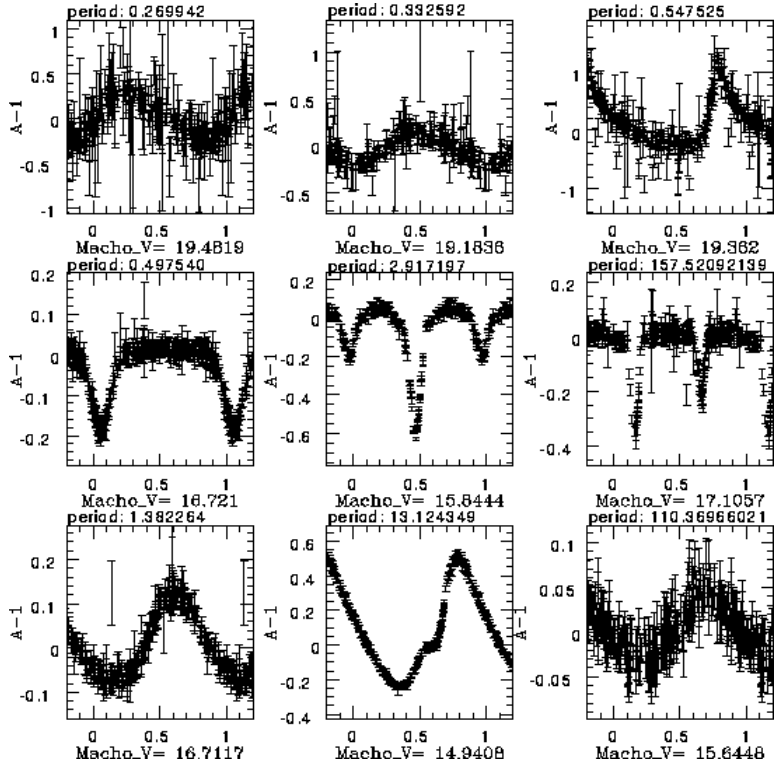

Figure 2. This figure shows phased MACHO_V light curves in flux units relative to the mean flux. The top row shows examples from the three peaks in the RR Lyrae period histogram. The second row shows a range of periods for eclipsing binaries. The third row shows an overtone and a fundamental mode Cepheid and a low amplitude Long Period Variable.

A preliminary analysis of this data provides the following census:

- ∼ 8000 RR Lyrae
- ∼ 1500 Cepheids
- ∼ 1200 eclipsing binaries from normal variable cuts
- ∼ 1000 additional eclipsing binaries from special analysis
- ∼ 19000 red variables

Figure 2 shows typical time series and folded light curves for each of these classes. It should be noted that since we observe the LMC throughout the year, and potential observing start times are dependent upon weather as well as the LMC's altitude, there is no strong aliasing and we are able to sample all phases of long and short period variables. In the following sections, I will briefly discuss each of these categories.

### 3.1. Cepheids

A clear indication that the MACHO database has much to offer in the study of variable stars came when a preliminary period-luminosity relation was plotted for periodic variables with the colors and magnitudes appropriate for Cepheid variables. This plot (Figure 3) makes clear the abundance of first overtone pulsators in the LMC—roughly 1/3 of the Cepheids are 1st overtone pulsators. More than 50% of these Cepheids are new identifications. It should be noted that much of the scatter in this plot is due to the fact that the 22 fields from which these light curves were drawn have not yet had a photometric zero point determined. A more complete discussion of work on Cepheids in the MACHO database can be found in this volume in the review by Welch et al. (1995).

### 3.2. RR Lyrae

RR Lyrae are faint enough that the typical measurement error of 10% is due to the mean seeing of 2.2 arcsec in this data set and the crowding in the bar of the LMC. Yet, periods are easily determined, and we have identified almost 8000 in 10.5 square degrees in the center of the LMC. Examination of the candidate variables from the eclipsing binary search has revealed additional, low amplitude and high average error RR Lyrae light curves, but we estimate that these 8000 represent at least 80% of the underlying population. The central density of RR Lyrae in this sample is about 1100 per square degree. This is significantly higher than estimates made by extrapolating from outer field densities assuming an exponential disk distribution (Kinman et al. 1991). This high central density may signal the presence of a PopII halo in the LMC. An histogram of the periods for this set of RR Lyrae is shown in Figure 4. Three peaks are seen at periods of 0.58, 0.34 and 0.28 days. The first two peaks represent RRab and RRc type stars which are fundamental and first overtone oscillators respectively. The peak at 0.28 days may represent second overtone oscillation. Light curves of these stars have small amplitudes but are more asymmetric than RRc stars. We would designate these RRe stars. The RRab have a broad period distribution and represent ∼ 80% of the RR Lyrae. This set RR Lyrae represents a powerful probe of the older, more metal poor population of the LMC as well as the physics of stellar pulsation.

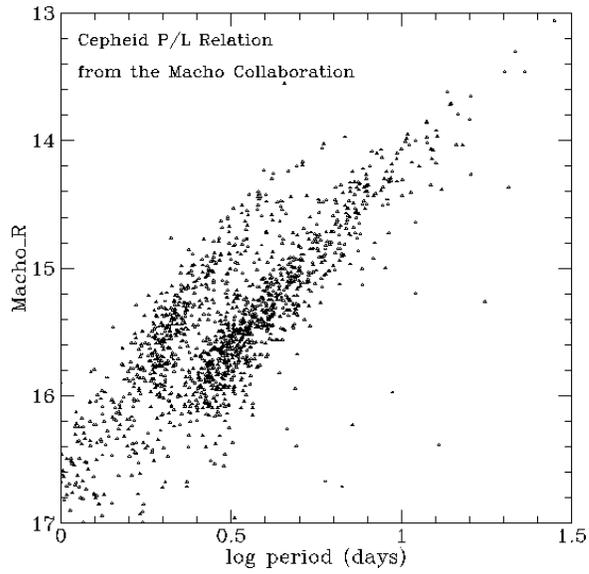

Figure 3. P/L relation for 1540 Cepheids from the first year's analysis of MACHO data.

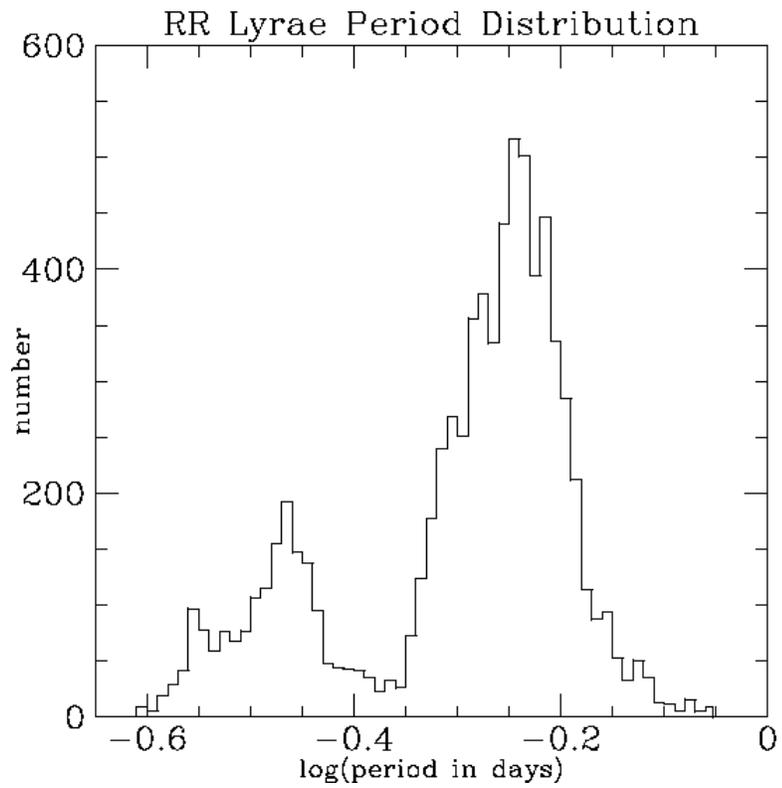

Figure 4. Histogram of periods found for 7652 RR Lyrae in the central 10.5 square degrees of the LMC.

### 3.3. Eclipsing Binaries

A great deal of stellar physics can be probed by study of the large number of eclipsing binaries found in the MACHO database. Much of this work will require that the data be placed upon a consistent photometric system so that model atmosphere calculations can be used to model observed light curves. It will also be necessary to determine our detection efficiency for eclipsing binaries with a large range of characteristics so that we can make statistically meaningful statements about the distribution of separations and eccentricities found during star formation in the LMC.

### 3.4. Red Variables

The bright, large amplitude, long period variables of the LMC have been well studied. The MACHO data on these stars provide unusual time resolution which may allow these variables to be more easily studied for the presence of chaotic behavior. The MACHO data will also yield direct, statistically meaningful, measures of the variability of such stars as M giants and carbon stars. Low amplitude, red variables are not well studied or understood. Overtone pulsations in the extended atmospheres of late type giants are expected to produce low amplitude light variations. Wood (this volume) points out that there is an unresolved question about the pulsation mode of Mira variables which would be answered by the identification and study of a first overtone Mira sequence. This seems quite likely because many very red, low amplitude, periodic variables have been found in the MACHO data. The last panel in Figure 2 shows the light curve of one such variable.

## 4. Aperiodic and Transient Variables in the LMC Data

Transient brightenings of stars are of particular interest to our collaboration since they may represent a potential source of confusion with microlensing. Our microlensing analysis has, in fact, identified a type of variable star which has a constant luminosity for extended periods of time, but shows occasional outbursts of 10-30% which we have termed 'bumps'. These bumpers are bright, main sequence stars with V $\sim 15 - 17$. These stars most often brighten more in our red passband than our blue passband. For more than half of these stars the episodes are clearly asymmetric and shorter than about 50 days with a more rapid brightening than dimming (Figure 5, top). The rest show more symmetric, and longer episodes (Figure 5, bottom). We have obtained spectra of seven of these stars and two of them showed strong H$\alpha$ and H$\beta$ emission while the rest showed evidence for filled cores in the Balmer lines. These stars are probably related to, or may be identical to, galactic Be stars. Percy and co-workers (1988, 1992) have noted possible outburst of a similar nature in galactic Be stars. Essentially all of the bumpers detected in our first year's data have shown further outbursts in our subsequent year's data. The apparent constancy of these stars between bumps argues that they are not a low amplitude pulsator.

Our data system is now reducing and analyzing images as they are accumulated, resulting in alerts being generated when a source deviates significantly from its previous behavior. This will allow us to alert the world to interesting transient stellar phenomena as well as to the beginning of microlensing events. If this system had been operational in November of 1992, we would have alerted the world to an LMC nova about two days earlier than it was first noticed, because this was our first data point during its rise. This capability will be used to identify bumpers in outburst for further spectroscopic study.

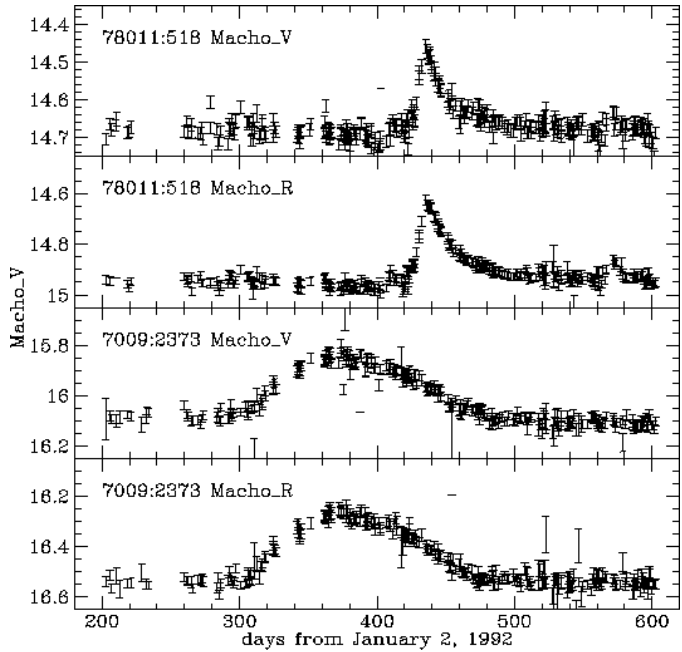

Figure 5. Light curves demonstrating the two general types of outbursts observed in the bumpers.

## 5. Bulge Variables

We have identified roughly 50,000 variables in our bulge data. Their light curves have been searched for periodicity in the same manner as the LMC data. The results of this phasing have not been analyzed at this time. It is possible to note that there are few, if any, bumpers in the bulge data, since these appear in the analysis for microlensing as well as variable stars.

## 6. Future Work

We are in the process of determining accurate magnitudes and transformations of our photometry to Cousins' V and R. Analysis of the bulge variables is just beginning and more LMC fields and bulge fields are currently being analyzed. There are currently more than 30,000 dual-color, 0.5 square degree images in our database, representing greater than 2 Tbytes of data. The photometry database is larger than 120 Gbytes and rapidly growing. We are currently reducing data in 'real' time allowing us to recognize microlensing events as the begin, and allowing us to detect the beginning of transient stellar phenomena as well. Perhaps the most exciting prospect is that our ability to alert the world to microlensing may allow concerted follow-up efforts to find details in microlensing curves indicating the existence of planets around low mass stars acting as lenses.

**Acknowledgments.** Work performed at LLNL is supported by the DOE under contract W-7405-ENG-48. Work performed by the Center for Particle Astrophysics on the UC campuses is supported in part by the Office of Science and Technology Centers of NSF under cooperative agreement AST-8809616. Work performed at MSSSO is supported by the Bilateral Science and Technology Program of the Australian Department of Industry, Technology and Regional Development.